\documentclass[11pt]{article}

\usepackage{aas_macros,amsmath,amssymb,comment,cite,esint,graphicx,mathtools}
\usepackage[margin=.8in,letterpaper]{geometry}
\usepackage[colorlinks=true]{hyperref}
\usepackage[affil-it]{authblk}
\usepackage{subcaption}
\usepackage[utf8]{inputenc}
\usepackage{mathrsfs}
\usepackage{appendix}
\usepackage{amssymb}
\usepackage{float}
\usepackage{color}
\usepackage{cite}
\usepackage{hyperref}
\hypersetup{pageanchor=false}
\usepackage{indentfirst}
\usepackage{url}
\usepackage{float}
\usepackage{caption}
\usepackage[numbers,square,comma,sort&compress,merge]{natbib}
\usepackage{esint}
\usepackage{overpic}
\usepackage{graphicx}
\usepackage{epsf,amsmath,bbold,amsfonts,stmaryrd}
\usepackage{textcomp}
\usepackage{ulem}
\usepackage{tikz}

\numberwithin{equation}{section}
\let\originalleft\left
\let\originalright\right
\renewcommand{\left}{\mathopen{}\mathclose\bgroup\originalleft}
\renewcommand{\right}{\aftergroup\egroup\originalright}
\mathcode`\*="8000
{\catcode`\*=\active\gdef*{\mathclose{}\,\mathopen{}}}

\newcommand{\be}{\begin{equation}}
\newcommand{\ee}{\end{equation}}
\newcommand{\bea}{\setlength\arraycolsep{2pt} \begin{eqnarray}}
\newcommand{\eea}{\end{eqnarray}}
\newcommand{\nn}{\nonumber}

\newcommand{\mM}{{\mathcal M}}

\def\a{\alpha}

\def\D{\Delta}
\def\f{\frac}

\def\lm{\lambda}

\def\m{\mu} 
\def\n{\nu} 
\def\nn{\nonumber}
\def\pl{\partial}
\def\p{\phi}

\def\t{\theta}

\def\be{\begin{equation}}
\def\ee{\end{equation}}
\def\bag{\begin{aligned}}
\def\eag{\end{aligned}}
\def\bea{\begin{eqnarray}}
\def\eea{\end{eqnarray}}
\def\ba{\begin{array}}
\def\ea{\end{array}}

\def\bc{\begin{center}}
\def\ec{\end{center}}

\begin{document}
\title{Observational signatures of rotating black holes in the semiclassical gravity with trace anomaly}

\author{Zhenyu Zhang$^{1}$, Yehui Hou$^{1\ast}$, Minyong Guo$^{2}$}
\date{}

\maketitle

\vspace{-10mm}

\begin{center}
{\it
$^1$Department of Physics, Peking University, No.5 Yiheyuan Rd, Beijing
100871, P.R. China\\\vspace{4mm}

$^2$ Department of Physics, Beijing Normal University,
No.19, Xinjiekouwai St, Beijing 100875, P. R. China\\\vspace{4mm}


}
\end{center}

\vspace{8mm}

\begin{abstract}

In a recent work by Fernandes \cite{Fernandes:2023vux}, an exact stationary and axisymmetric solution was discovered in semiclassical gravity with type-A trace anomaly, identified as a quantum-corrected version of the Kerr black hole. In this study, we explore the observational signatures of this black hole solution. Our investigation reveals that there exist prograde and retrograde light rings, whose radii increase monotonically with the coupling parameter $\alpha$. When $\alpha$ is negative, the shadow area for the quantum-corrected black hole is smaller than that of the Kerr black hole, whereas when $\alpha$ is positive, the area is larger. For a near-extremal black hole, its high-spin feature (the NHEKline) is found to be highly susceptible to disruption by $\alpha$. Furthermore, we discuss the images of the quantum-corrected black hole in the presence of a thin accretion disk and compare them to those of the Kerr black hole.   Our study highlights the importance of near-horizon emission sources in detecting the effects of quantum corrections by black hole images.
\end{abstract}

\vfill{\footnotesize $\ast$ Corresponding author: yehuihou@pku.edu.cn}

\maketitle

\newpage
\baselineskip 18pt
\section{Introduction}\label{sec1}

Semiclassical gravity is an approach that considers the backreaction of quantum fields while treating spacetime classically. One of the quantum effects in this scheme is the trace anomaly, which refers to the breaking of symmetry in a conformally invariant classical theory due to one-loop quantum corrections \cite{Capper:1974ic}. As a result of the trace anomaly, the renormalized stress tensor of quantum fields has a non-zero trace, serving as a source term in the semiclassical Einstein equations. The trace anomaly may also induce higher-order curvature terms such as the Gauss-Bonnet term, which arises from the type-A anomaly \cite{Deser:1993yx}. 

One of the topics of solutions in semiclassical gravity is the corrected versions of classical black holes that account for quantum effects. However, the derivations are challenging because the renormalized stress tensor is often unknown, requiring additional assumptions to solve the problem. More than a decade ago, considering only the type-A anomaly, \cite{Cai:2009ua} was the first to find the static and spherically symmetric black hole solution in a four-dimensional spacetime within the framework of such semiclassical gravity. This spherically symmetric black hole has been widely studied in subsequent works \cite{Guo:2020zmf,Zeng:2020dco,Islam:2020xmy,Konoplya:2020bxa,Tsujikawa:2023egy} because of its intriguing features resulting from quantum effects. Furthermore, having an exact stationary and axisymmetric solution to the semiclassical Einstein equations that is sourced by the type-A trace anomaly is crucial for modeling the actual black hole in space.

Very recently, by adopting a Kerr-Schild ansatz with stationary and axisymmetric configuration, the author in \cite{Fernandes:2023vux} analytically solved the semiclassical Einstein equation and obtained a Kerr black hole solution with the type-A trace anomaly. Compared to the classical Kerr black hole, this new solution replaces the ADM mass with a mass function
 \be\label{MM}
\mathcal{M}(r,\t) = \f{2 M}{1 + \sqrt{1-\f{8 \a r \xi M}{\Sigma^3}}} \, 
 \ee
in the ingoing Kerr-like coordinates, where $M$ represents the ADM mass, $\Sigma=r^2+a^2\cos^2\t$ with $a$ denoting the spin parameter, $\xi=r^2-3a^2\cos^2\t$, and $\a$ representing the coupling constant of the type-A anomaly. This rotating black hole solution includes quantum corrections and reduces to the classical Kerr spacetime when $\a=0$. Furthermore, when $a=0$, the solution reduces to the static and spherically symmetric solution in semiclassical gravity. This new solution presents several unique characteristics. For instance, the event horizon geometry is non-spherically symmetric, and there exists another Killing horizon outside of it. Additionally, under specific coupling constants, the spin parameter may surpass the traditional Kerr bound. This suggests that black holes may possess higher spins than their classical counterparts \cite{Jiang:2023gpx}.

From an astrophysical perspective, the rotating black hole with the type-A anomaly may have some observational features. With the Event Horizon Telescope collaboration already capturing images of supermassive black holes at the centers of galaxies \cite{EventHorizonTelescope:2019dse,EventHorizonTelescope:2022wkp,Lu:2023bbn}, studying black hole images, particularly the shadow of this black hole, becomes essential \cite{Vagnozzi:2022moj, Chen:2022scf}. The size and shape of a shadow can reflect the geometric structure and physical properties of the central black hole, thus having the potential to test the coupling parameter of the quantum-corrected black hole. 

In this study, we investigate the observable features of the recently discovered black hole \cite{Fernandes:2023vux}. Specifically, our focus lies on the analysis of light rings (LRs) \cite{Cunha:2017qtt,Cunha:2020azh,Guo:2020qwk,Ghosh:2021txu}, shadows \cite{Heydari-Fard:2021ljh, Peng:2020wun, Zhang:2022osx, Meng:2022kjs, He:2022opa}, and images when illuminated by an external light source \cite{Cunha:2019hzj, Zhang:2021hit, Hou:2022gge, Hou:2022eev, Wang:2023vcv, Zhang:2023okw, Hu:2023bzy}. Initially, we examine the effective potential of particles in the equatorial plane and derive the equation governing the positions of the LRs. Due to the absence of an analytical solution, we numerically calculate the LRs as a function of the coupling constant for various spin values. The existence of LRs implies the presence of a critical curve and a shadow on the observer's screen when illuminated by celestial light. To explore the shadow images, we employ the backward ray-tracing method. Furthermore, we utilize a simplified thin disk model to compare the intensity images with those of the Kerr black hole, aiming for a more realistic scenario.

The paper is organized as follows. In Sec. \ref{sec2}, we provide a review of the quantum-corrected Kerr black hole, examining particle trajectories and the existence of light rings within this spacetime. In Sec. \ref{sec3}, we investigate the shadow images of the quantum-corrected Kerr black hole when illuminated by a celestial source. Then, in Sec. \ref{sec4}, we introduce a thin disk model that serves as a more realistic light source, presenting the obtained results of black hole images. Finally, we summarize our findings and conclude this study in Sec. \ref{sec5}. We work in the geometrized unit with $8\pi G = c = 1$ in this paper.

\section{The metric and the light rings}\label{sec2}

\begin{figure}[h!]
	\centering
	{\includegraphics[scale=0.9]{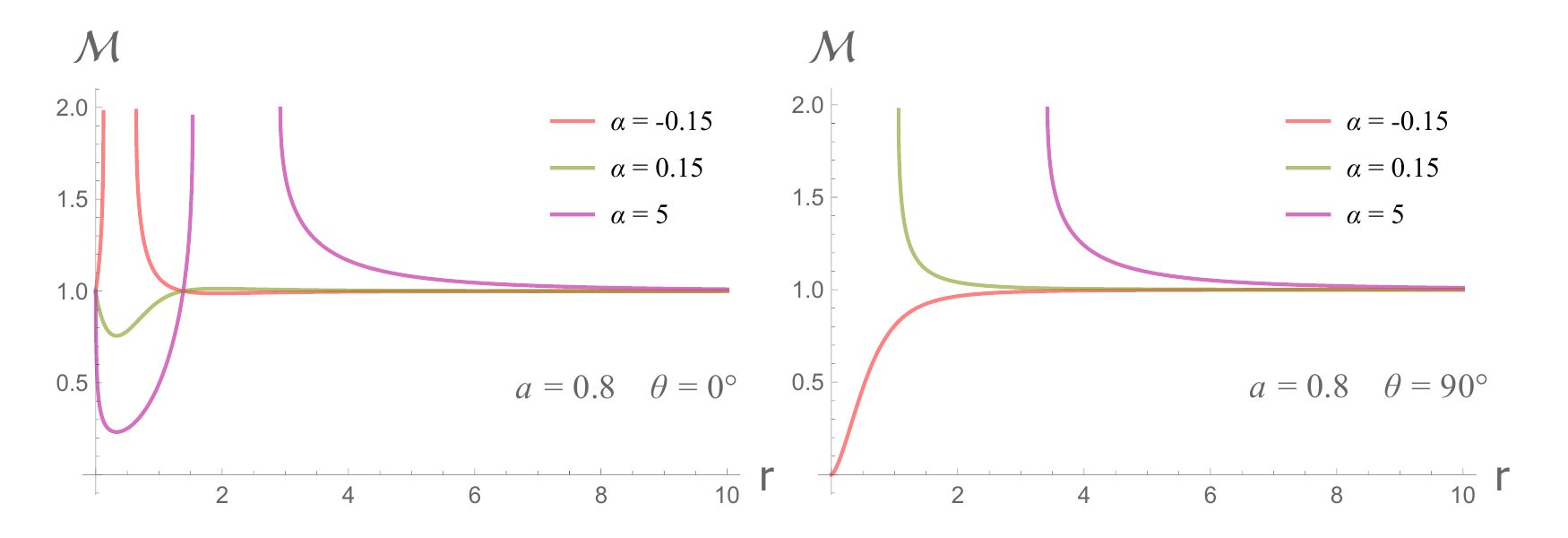}}	
	\caption{$\mathcal{M}$ as function of $r$ for differnet values of $\a$, evaluated at the pole (Left) and the equatorial plane (Right). The spin is fix at $a = 0.8$.}
	\label{M}
\end{figure}

To begin, let us briefly review the semiclassical Einstein gravity and the quantum-corrected Kerr black holes in \cite{Fernandes:2023vux}. In this framework, the background geometry remains classical while the quantum fields influence the geometry through their expectation value of the renormalized stress tensor, denoted as $\langle T_{\m\n} \rangle$, in the Einstein equations. Due to the quantum anomaly, the trace of $\langle T_{\m\n} \rangle$ is non-zero and dependent only on the local curvature. If we only consider the type-A anomaly, we have
\be\label{ta}
 g^{\m\n}\langle T_{\m\n} \rangle = \f{\a}{2}\, \mathcal{G} \, ,
\ee
where $\mathcal{G} = R^2-4R_{\m\n}R^{\m\n} + R_{\m\n\rho\sigma} R^{\m\n\rho\sigma}$ is the Gauss-Bonnet scalar, and $\a$ represents the coupling constant. By combining Eq.~\eqref{ta} with the semiclassical Einstein equations, $R_{\m\n} - \f{1}{2}g_{\m\n}R = \langle T_{\m\n} \rangle$, one arrives at:
\be\label{RG}
R = \f{\a}{2}\mathcal{G} \, .
\ee
In general, it is impossible to solve the semiclassical Einstein equations when the renormalized stress tensor remains undetermined. However, by adopting a Kerr-Schild ansatz and directly solving Eq.~\eqref{RG}, the author of \cite{Fernandes:2023vux} found a stationary and axisymmetric solution, which is interpreted as a quantum-corrected Kerr black hole. The line-element was written in the ingoing Kerr-like coordinates $(\nu, r, \t, \varphi)$, 
\be\label{KS}
\mathrm{d}s^2=-\left(1-\frac{2\mathcal{M}r}{\Sigma}\right)\left(\mathrm{d}\n-a\sin^2{\t}\mathrm{d}\varphi\right)^2+ 2\left(\mathrm{d}\n - a \sin^2{\t} \mathrm{d}\varphi\right)\left(\mathrm{d}r-a\sin^2{\t}\mathrm{d}\varphi\right) + \Sigma\left(\mathrm{d}\theta^2+\sin^2{\t}\mathrm{d}\phi^2\right)\,,
\ee
where $\mathcal{M}$ is the mass function in Eq.~\eqref{MM}. Since $\mathcal{M}$ is dependent on $\t$, the resulting spacetime does not satisfy the circularity conditions \cite{Delaporte_2022}. Fig.~\ref{M} provides various examples of $\mathcal{M}$ under typical coupling constants. It is worth noting that a significant difference exists between the cases with positive and negative coupling constants. Moreover, the radius of the event horizon is also a function of $\t$, rendering it non-spherically symmetric and requiring numerical solving. Besides, it has been observed that another Killing horizon is present at $r^2-2\mathcal{M} r + a^2 = 0$, which does not coincide with the event horizon but is located quite close to it.

In this work, we focus on potential observational features of the quantum-corrected Kerr black hole. Specifically, we would like to solve the propagations of light and determine the associated black hole images. Under the WKB approximation, i.e., the wave length of the light is much smaller than the gravitational radius, the light propagations can be described by null geodesics. The constrained Hamiltonian for geodesics is given by 
\be\label{H}
\mathcal{H} = \f{1}{2} g_{\m\n} p^{\m} p^{\n}  = -\f{1}{2}m^2 \, ,
\ee
where $p^{\m}$ is the four momentum, $m = 0, 1$ denotes null and timelike geodesics, respectively. The equations of motion are
\bea
\f{\pl \mathcal{H}}{\pl p_\m} = \dot{x}^\m \, ,  \quad \f{\pl \mathcal{H}}{\pl x^\m} = -\dot{p}_\m \, ,
\eea
where the dot denotes the derivative with respect to the affine parameter. As the spacetime is non-circular, it seems impossible to recast the equations of motion into first-order equations as can be done in Kerr spacetime. Moreover, the global transformation between the ingoing Kerr coordinates and the Boyer-Lindquist coordinates is rather complicated \cite{Zi:2023qfk}.  

Nonetheless, we can still gain some insights from the simplified scenario of geodesics on the equatorial plane. When restricted to $\t = \pi/2$, the metric Eq.~\eqref{KS} can be easily transformed to 
\bea
\mathrm{d}\nu=\mathrm{d}t + \f{r^2-a^2}{\D}\mathrm{d}r \, , \quad \mathrm{d}\varphi =\mathrm{d}\phi + \f{a}{\D} \mathrm{d}r \, ,
\eea
\be
\mathrm{d}s^2=-\left(1-\frac{2\mathcal{M}_0r}{\Sigma}\right)\mathrm{d}t^2+\frac{\Sigma}{\Delta}\mathrm{d}r^2+\left(r^2+a^2+\frac{2\mathcal{M}_0ra^2}{\Sigma}\right)\mathrm{d}\phi^2-\frac{4\mathcal{M}_0ra}{\Sigma}\mathrm{d}t\mathrm{d}\phi\,,
\label{BL}
\ee
where $\Delta=r^2-2\mathcal{M}_0r+a^2, \mathcal{M}_0 =\mathcal{M}(r,\pi/2)= 2 M /\left( 1 + \sqrt{1-\f{8 \a M}{r^3}} \right)$. 
In this case, the geodesics are determined by an one-dimensional equation of $r$, which takes
\bea \label{Veff}
\bigg( \f{\mathrm{d}r}{\mathrm{d}\lm} \bigg)^2 = V_{eff}(r) = E^2 -m^2 +\f{2\mathcal{M}_0m^2}{r} + \f{a^2(E^2-m^2)-L^2}{r^2} + \f{2\mathcal{M}_0(L-a E)^2 }{r^3}
\eea   
with  $E = -p_t, L = p_{\p}$ the conserved energy and angular momentum, and $V_{eff}$ acting as an effective potential. Let us now focus on the case of photons with  $m = 0$. The key property of photons is that their conserved energy can be absorbed into the affine parameter by rescaling $\lm \rightarrow  \lm E$. Then, by defining $l \equiv L/E$, the potential is simplified to
\be
V_{eff}(r) = 1 + \f{a^2-l^2}{r^2} + \f{2\mathcal{M}_0(l-a )^2 }{r^3} \, .
\ee
The unstable photon orbits, known as light rings (LRs), are circular orbits of photons that play a crucial role in the observational properties of black holes. LRs can be obtained by setting
\be
V_{eff} = 0 \, , \quad  \pl_r V_{eff} = 0 \, .
\ee
From these equations, we obtain
\bea\label{ll}
&& l = -\f{a\bigg(r^3+9r-8\a+6r^2\sqrt{1-\f{8\a}{r^3}}\,\bigg)}{r^3-9r+8\a} \, , \nn \\
&& 1 + \f{a^2-l^2}{r^2} + \f{4(l-a)^2}{r^3\sqrt{1-\f{8\a}{r^3}}} = 0 \, ,
\eea
where and thereafter, we set $M=1$, without loss of generality. Eq.~\eqref{ll} provides the equation for the LRs,
\be\label{elr}
1 +\f{16a^2\bigg(r^3-8\a+3r^2\sqrt{1-\f{8\a}{r^3}} \, \bigg)^2}{r^3\big(r^3-9r-8\a\big)^2\bigg(1+\sqrt{1-\f{8\a}{r^3}}\, \bigg)} + \f{a^2}{r^2} - \f{a^2\bigg(r^3+9r-8\a+6r^2\sqrt{1-\f{8\a}{r^3}}\,\bigg)^2}{r^2(r^3-9r-8\a)^2} = 0 \, .
\ee

\begin{figure}[h!]
	\centering
	{\includegraphics[scale=0.085]{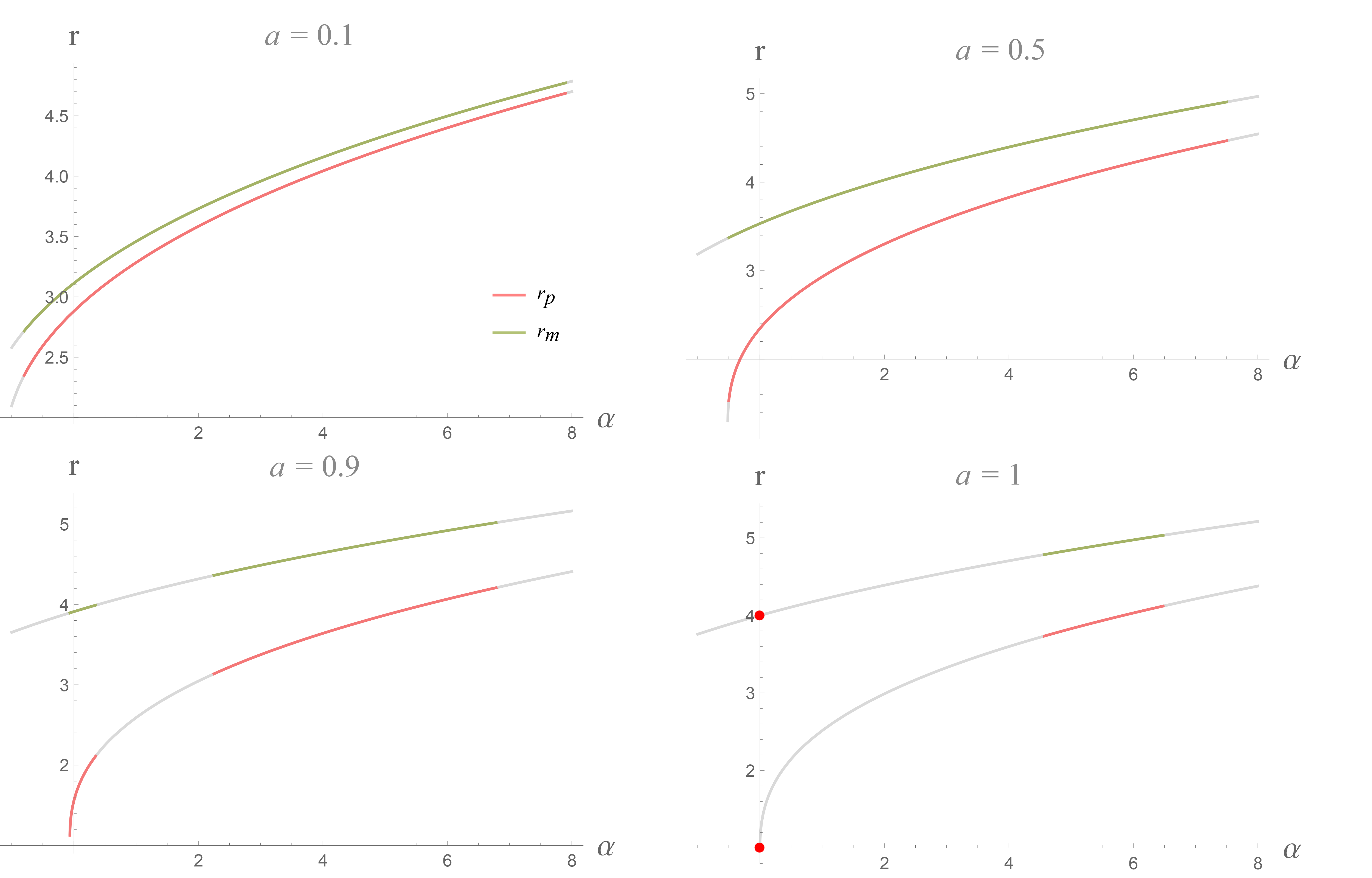}}
	\caption{LRs as functions of $\a$, under different spin parameters. The red and orange colors denote the prograde orbit and retrograde orbit, respectively. }
	\label{LRs}
\end{figure}

As Eq.~\eqref{elr} is very complicated, we solve the LRs numerically. There are two real roots, $r_{p}$, $r_m$, corresponding to prograde and retrograde photon orbits respectively. We have confirmed that these LRs are unstable under radial perturbations, that is, $\pl^2_r V_{eff} > 0$. Consequently, similar to the Kerr spacetime, the LRs in the quantum-corrected Kerr solution imply 
a $shadow$ region in the image of such black hole.

Fig.~\ref{LRs} illustrates how the radii of LRs change as the coupling constant $\a$ increases, while the spin parameter remains fixed at different values. It's worth noting that although Eq.~\eqref{elr} has solutions throughout the range of $ -1\leq\a \leq 8$, only the colored parts of the curves correspond to the actual black hole solution, i.e., where the event horizon encloses the curvature singularities. The gray portions of the curves correspond either to cases where the event horizon is complex or where the singularity crosses the event horizon. For the case of $a=0.9$ (bottom right), there are two distinct intervals for black hole solutions: $-0.0649\leq\a\leq0.3487$ and $2.25\leq\a\leq6.784$, consistent with the domain of black hole solutions found in \cite{Fernandes:2023vux}. In the case of $a=1$ (bottom left), the black hole touches the Kerr bound, and a black hole still exists for $\a = 0$ (red dots) and $4.5698\leq\a\leq6.47976$.  Anyway, it is apparent that both
$r_p$ and $r_m$ increase with $\a$. This implies that a larger value of $\a$ could potentially result in a larger 
shadow in the black hole image. This finding aligns with previous research studies \cite{Guo:2020zmf,Zeng:2020dco,Kumar:2020owy}.

\section{Shadows cast by a celestial light source}\label{sec3}

It is known that the LRs suggest the existence of a family of unstable photon orbits around the black hole, which significantly affect the observational characteristics of the black hole. Photons launched near these orbits complete multiple loops around the black hole before reaching the observer, leading to the formation of a critical curve in the observer's screen \cite{Gralla:2019xty}. To investigate these characteristics of the quantum-corrected Kerr black hole, we utilize a celestial sphere as the light source to illuminate the black hole \cite{Hu:2020usx}. The black hole is located at the origin, whose size is much smaller than that of the celestial sphere and the distance between the observer and the origin. This setup allows the celestial sphere to outline the critical curve accurately and reveal the unstable photon orbits. When illuminated by the celestial sphere, any spacetime information that lies behind the LRs remains invisible on the screen. Light rays that penetrate the interior of the prograde LR $r_p$ are inevitably captured by the black hole, creating a shadow region on the screen.

Once the celestial sphere model is established, we utilize the backward ray-tracing method to generate the black hole images. The numerical strategy involves setting up a camera model at the observer and integrating the equations of motion along the null geodesics moving backward from the observer. More specifically, we employ a fisheye camera that incorporates the stereographic projection of the momentum $p_\m$ of photons onto the screen. With the values of $p_\m$ on the screen determined, we may now proceed with solving the trajectories of photons in the metric Eq.~\eqref{KS} by performing backward integration of the Hamiltonian equations,
\bea
\f{\pl \mathcal{H}}{\pl p_\m} = \dot{x}^\m \, ,  \quad \f{\pl \mathcal{H}}{\pl x^\m} = -\dot{p}_\m \, ,
\eea
where the dot denotes the derivative concerning the affine parameter. During the ray-tracing process, light rays that reach the celestial sphere are colored based on their positions. Rays that reach the horizon are colored black, creating a shadow region on the screen.

\begin{figure}[h!]
	\centering
	\includegraphics[width=6in]{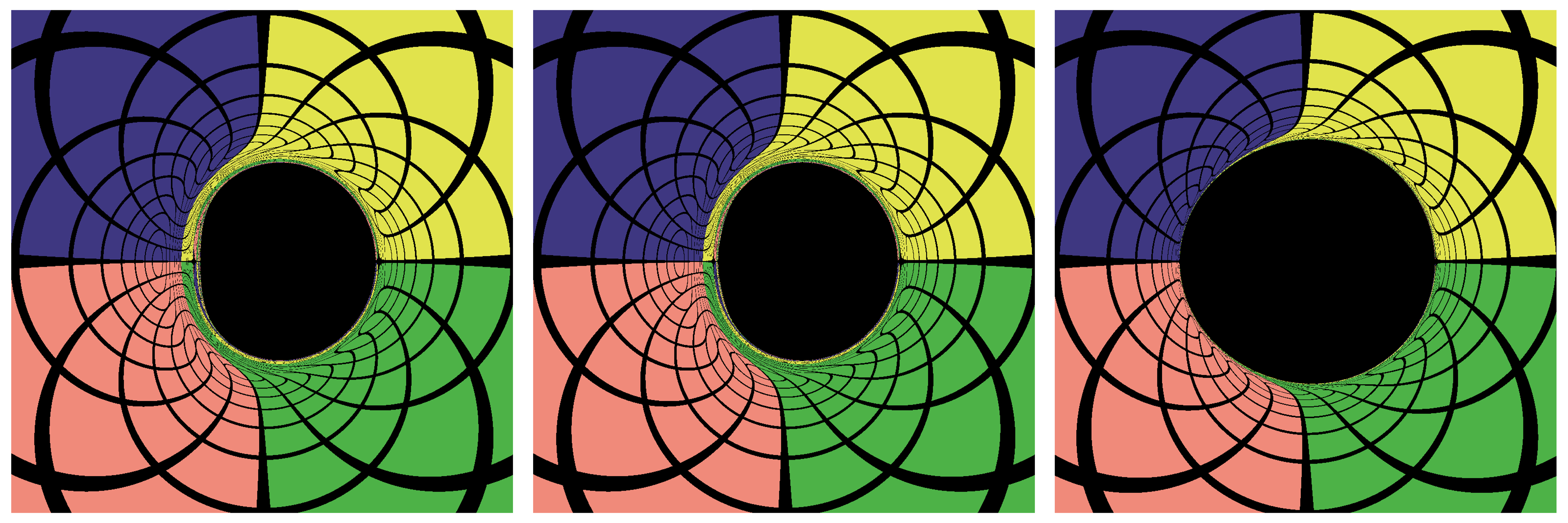}
	
	\caption{Black hole images of $a=0.99$ with $\a \geq 0$. Left: $\a=0$, which corresponds to a near extreme Kerr black hole. Middle: $\a=0.02$. Right: $\a=6.2$. The observer is placed at $r_o=300$, $\theta_o=\pi/2$.}
	\label{shadows099}
\end{figure}

\begin{figure}[h!]
	\centering
	{\includegraphics[scale=0.9]{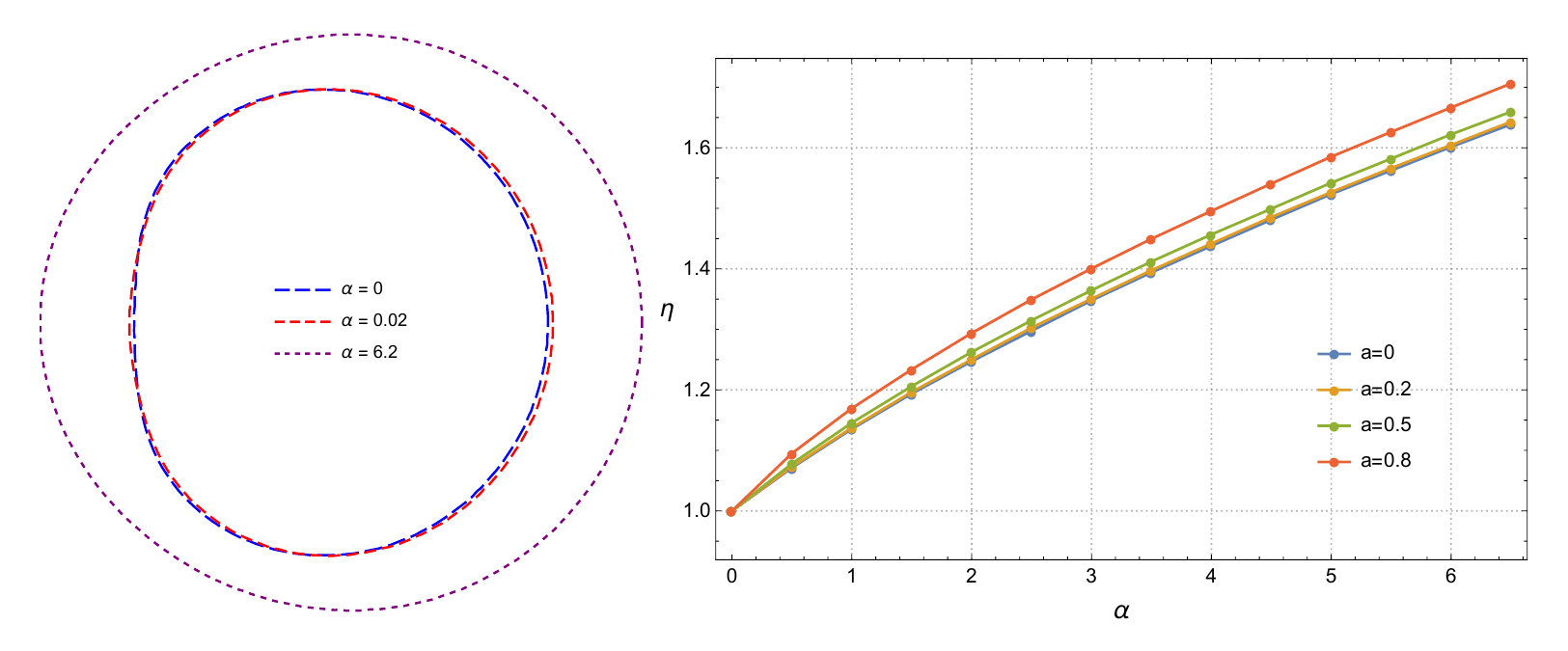}}
	\caption{Left: shadow curves of black holes with $a=0.99$ and $\alpha \geq 0$. Right: the variation of the area ratio $\eta$ with respect to postive $\a$ for different black hole spin $a$. The incline angle of the observer is fixed at $\theta_o=\pi/2$.}
	\label{area1}
\end{figure}

\begin{figure}[h!]
	\centering
	\includegraphics[width=6in]{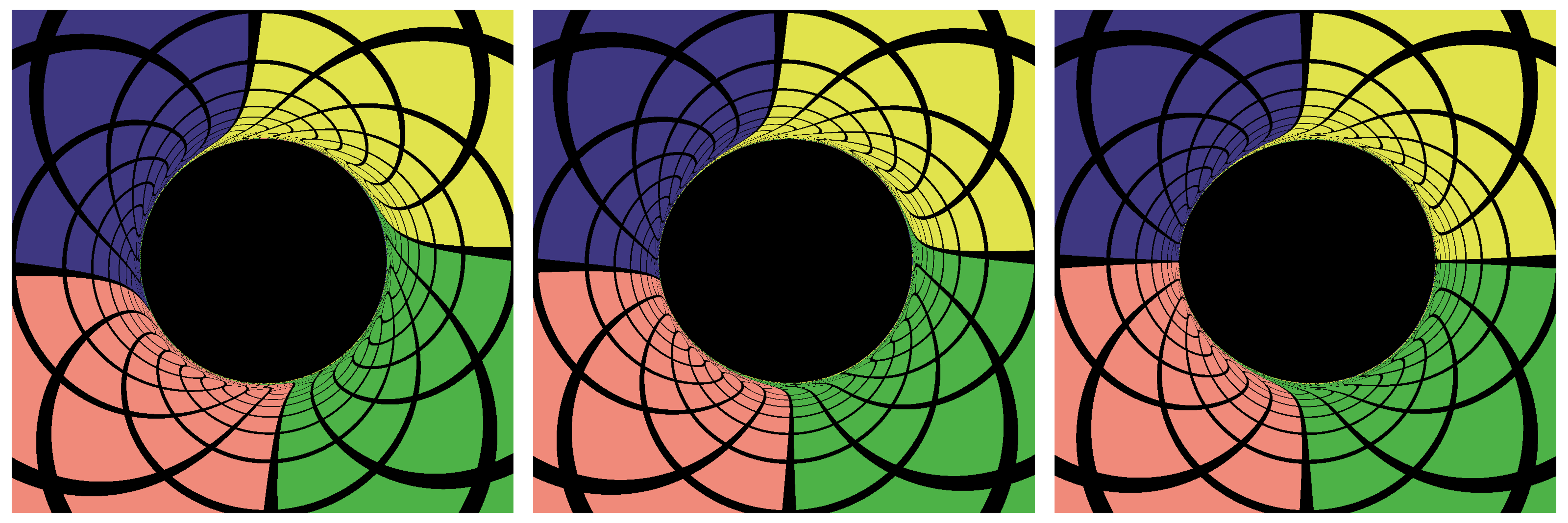}
	\caption{Black hole images of $a=1.06$ and $\a=6.275$, which violates the Kerr bound $a \leq 1$. The observer is placed at $r_o=300$. And the incline angles of the observer are $\t_o=\pi/9, \pi/3, \pi/2$, respectively.}
	\label{shadows106}
\end{figure}

Let us first consider the case of $\a \geq 0$. Fig.~\ref{shadows099} displays black hole images for three different positive values of  $\a$, with an identical black hole spin $a=0.99$ for an edge-on observer ($\t_o = \pi/2$). It is evident that the black hole shadow increases in size due to the impact of a positive $\a$. As $\a$ becomes sufficiently large, the shadow shape transforms into an ellipse as demonstrated in the right panel of Fig.~\ref{shadows099}. Moreover, regarding the case of large $\a$, the lensed images surrounding the shadow become narrower, indicating a different gravitational lensing process compared to that in Kerr spacetime. To quantify the impact of quantum corrections on the shadow, we introduce a parameter $\eta\equiv S_{\text{BH}}/S_{\text{Kerr}}$, representing the area ratio between the shadow of a quantum-corrected black hole and that of a Kerr black hole with identical spin. The right panel of Fig.~\ref{area1} shows the variation of $\eta$ concerning $\a$ under different black hole spins. The area ratio $\eta$ progressively increases as the quantum-corrected parameter grows, with a more substantial impact on higher-spin black holes. This finding aligns with the behavior of the LRs, whose radii are monotonically increasing functions of $\a$. Furthermore, as illustrated in Fig.~\ref{shadows106}, we present an example of a violation of the Kerr bound, where $a = 1.06$ and $\a = 6.275$, which is close to the maximum value found in \cite{Fernandes:2023vux}. However, we do not observe a remarkable feature compared to the right panel of Fig.~\ref{shadows099}, where $a=0.99$ and $\a = 6.2$. It is possible that the quantum corrections suppress the high-spin effect, even when it surpasses the Kerr bound.

\begin{figure}[h!]
	\centering
	\includegraphics[width=6in]{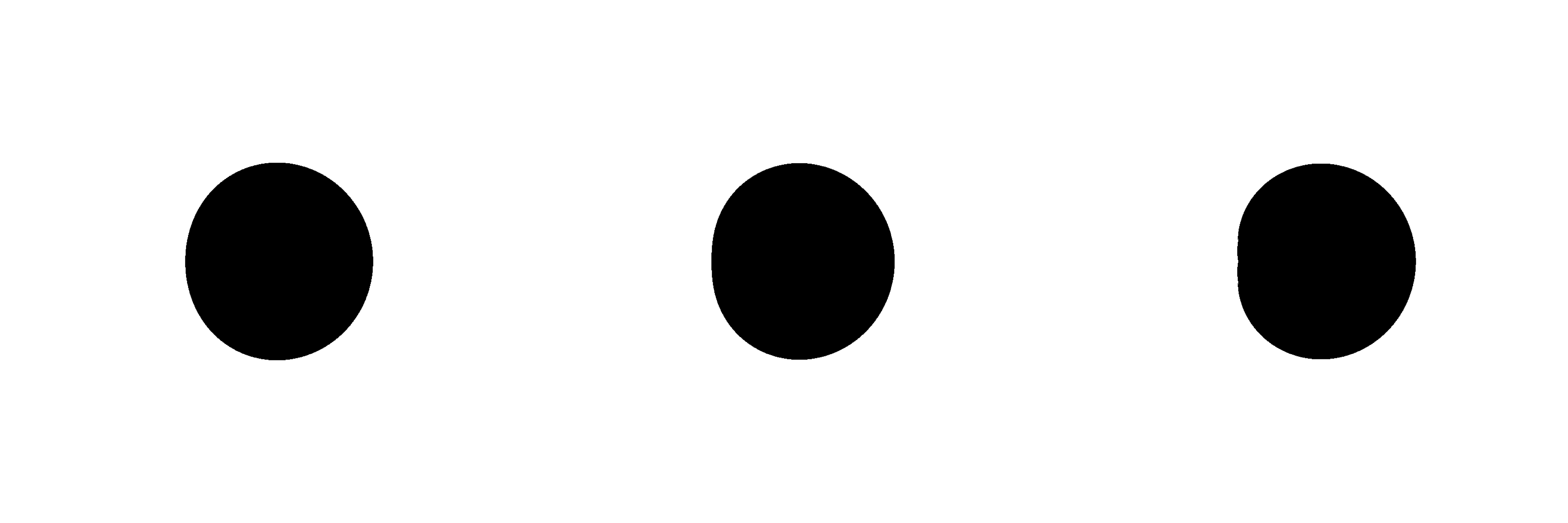}

	\caption{Black hole shadows of $a=0.8$ with $\alpha \leq 0$. Left: $\alpha=0$, which corresponds to a Kerr black hole. Middle: $\alpha=-0.09$. Right: $\alpha=-0.15$. The observer is placed at $r_o=300$, $\theta_o=\pi/2$.}
	\label{shadows08}
\end{figure}

\begin{figure}[h!]
	\centering
	{\includegraphics[scale=0.9]{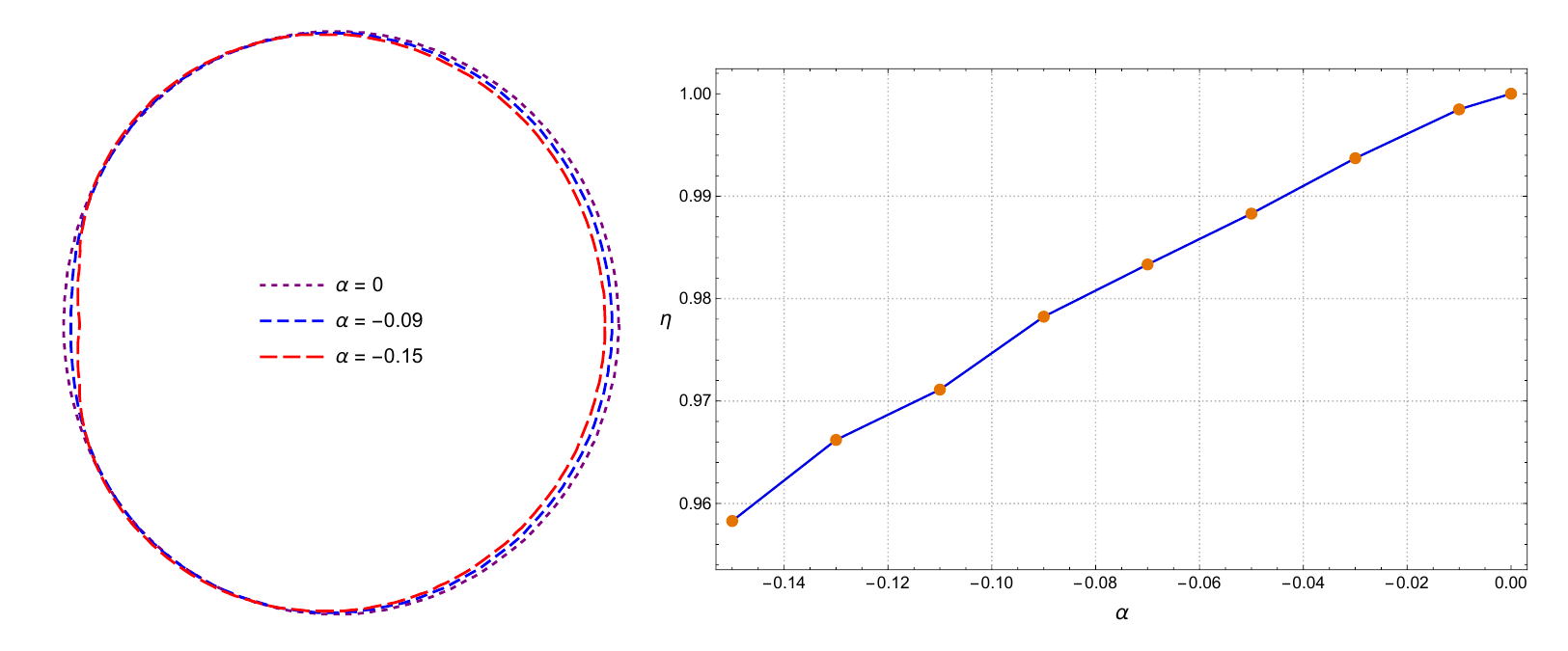}}
	\caption{Left: shadow curves of black holes with $a=0.8$ and $\alpha \leq 0$. Right: the variation of the area ratio $\eta$ with respect to negative $\alpha$. The black hole spin is fixed at $a=0.8$ and the incline angle of the observer is fixed at $\theta_o=\pi/2$.}
	\label{area2}
\end{figure}

Next, we proceed to the case of $\a \leq 0$. For clarity, we present the black hole shadows for three different values of $\a$, but with the same spin $a=0.8$ for an edge-on observer. The results are shown in Fig.~\ref{shadows08}. Here, we choose a smaller spin than that of the case with $\a \geq 0$, since the high spin limits the range of values for $\alpha$, as shown in Fig. 2 of \cite{Fernandes:2023vux}. As $\a$ decreases, the left-hand side of the shadow curve exhibits a concave shape, as seen in the right panel of Fig.~\ref{shadows08} and the left panel of Fig.~\ref{area2}. Due to such effect, the shadow area may decrease as $\a$ decreases. To verify this, we plot the variation of the area ratio $\eta$ concerning $\a$ in this case in Fig.~\ref{area2}. The range of $\a$ follows the domain of existence for a black hole. By combining Fig.~\ref{area2} with Fig.~\ref{area1}, we conclude that the area ratio $\eta$ is a monotonically increasing function concerning the quantum-corrected parameter $\a$, similar to the radii of LRs. 

Moreover, in the case of a nearly extreme Kerr black hole, an edge-on observer can observe a vertical line segment within the left contour of the shadow, as depicted in the left panel of Fig.\ref{shadows099}. This vertical line, known as the NHEKline \cite{Gralla:2017ufe}, originates from photons emitted in the near-horizon-extreme-Kerr (NHEK) geometry of the Kerr spacetime \cite{Bardeen:1999px, Gralla:2017ufe, Hou:2023hto}. The existence of the NHEK geometry is attributed to the degeneracy of the inner and outer horizons \cite{Bardeen:1999px}. However, the presence of the quantum-corrected parameter $\a$ breaks the global degeneracy of the horizons, thereby destroying the NHEK geometry. As demonstrated in the left panel of Fig.\ref{area1}, NHEKlines can be blurred by even a small value of $\a$.

\section{Black hole images with a thin accretion disk}\label{sec4}

The millimeter-wave images of supermassive black holes are typically thought to be dominated by accretion disks \cite{EventHorizonTelescope:2019dse, EventHorizonTelescope:2022wkp, Lu:2023bbn}. In this section, we use the disk model employed in \cite{Hou:2022eev} to generate images of the black hole-disk system. The disk is geometrically and optically thin, placed on the equatorial plane. For simplicity, we determine the disk structure in the Boyer-Lindquist coordinates Eq.~\eqref{BL}. The accretion flows are divided into two parts by the innermost stable circular orbit (ISCO), whose radius is determined by solving the following equations:
\bea\label{iscoeq}
V_{eff}|_{r=r_{\text{ISCO}}} = 0 \, , \quad  \pl_r V_{eff}|_{r=r_{\text{ISCO}}} = 0 \, , \quad  \pl^2_r V_{eff}|_{r=r_{\text{ISCO}}} = 0 \, ,
\eea
where $V_{eff}$ is given by Eq. (\ref{Veff}) with $m=1$. The above equations are quite complicated, and the radius of the ISCO can only be solved numerically. Outside the ISCO, the accretion flows move along timelike circular orbits,
\be
 u^\m = u^t_{\text{out}}\left( 1\, ,0\, ,0\, ,\Omega_s\right) \, ,
\ee
where
\bea
u^t_{\text{out}} = \sqrt{-\frac{1}{g_{\phi\phi}\Omega_s^2 + 2 g_{t\phi}\Omega_s+g_{tt}}} \,\, \Bigg|_{\t = \pi/2}  \, , \quad
\Omega_s = \frac{-\pl_r g_{t\phi} + \sqrt{\left(\pl_r g_{t\phi}\right)^2 -\pl_r g_{\phi\phi} \pl_r g_{tt}}}{\pl_r g_{\phi\phi}}\,\, \Bigg|_{\t = \pi/2} \,.
\eea
Inside the ISCO, the fluid falls from the ISCO to the event horizon, with the same conserved quantities as that for the ISCO. The components of the four-velocity take the form:
\bea
&&u^t_{\text{in}}= (- g^{tt} E_{\text{ISCO}} +  g^{t\phi} L_{\text{ISCO}}) \,\, \big|_{\t = \pi/2}\, , \quad
u^\phi_{\text{in}}= (- g^{t\phi} E_{\text{ISCO}}  + g^{\phi\phi} L_{\text{ISCO}} ) \,\, \big|_{\t = \pi/2}\, ,\nn \\
&&u^r_{\text{in}}=-\sqrt{-\frac{g_{tt} u^t_{\text{in}} u^t_{\text{in}} + 2 g_{t\phi} u^t_{\text{in}} u^\phi_{\text{in}} +g_{\phi\phi}u^\phi_{\text{in}} u^\phi_{\text{in}} +1}{g_{rr}}} \,\, \Bigg|_{\t = \pi/2} \, , \quad u^\theta_{\text{in}} = 0 \, ,
\eea
where $E_{\text{ISCO}}$ and $L_{\text{ISCO}}$ is the energy and angular momentum at the ISCO.

Light rays propagating around the black hole can undergo multiple crossings with the disk, leading to the accumulation of intensity. The expression for the observed intensity can be formulated as follows:
\be
I_{\nu_o} = \sum_{n=1}^{N_{max}} g^3_n J_n \, ,
\ee
where $\nu_o$ is the observed frequency on the screen, $n=1...N_{max}$ is the number of times that
the ray crosses the disk and $N_{max}$ is the maximal crossing number, $g_n$ is the corresponding redshift factor, and $J_n$ denotes the emissivity at the $n$-th intersection point. The emissivity is chosen as a function of $r$,
\be\label{ems}
J = \text{exp}\left( -\frac{1}{2}z^2 -2 z \right)\, , \quad z=\text{log}\frac{r}{r_H}\,,
\ee
where $r_H$ is the horizon radius on the equatorial plane. 

\begin{figure}[h!]
	\centering
	{\includegraphics[width=6.5in]{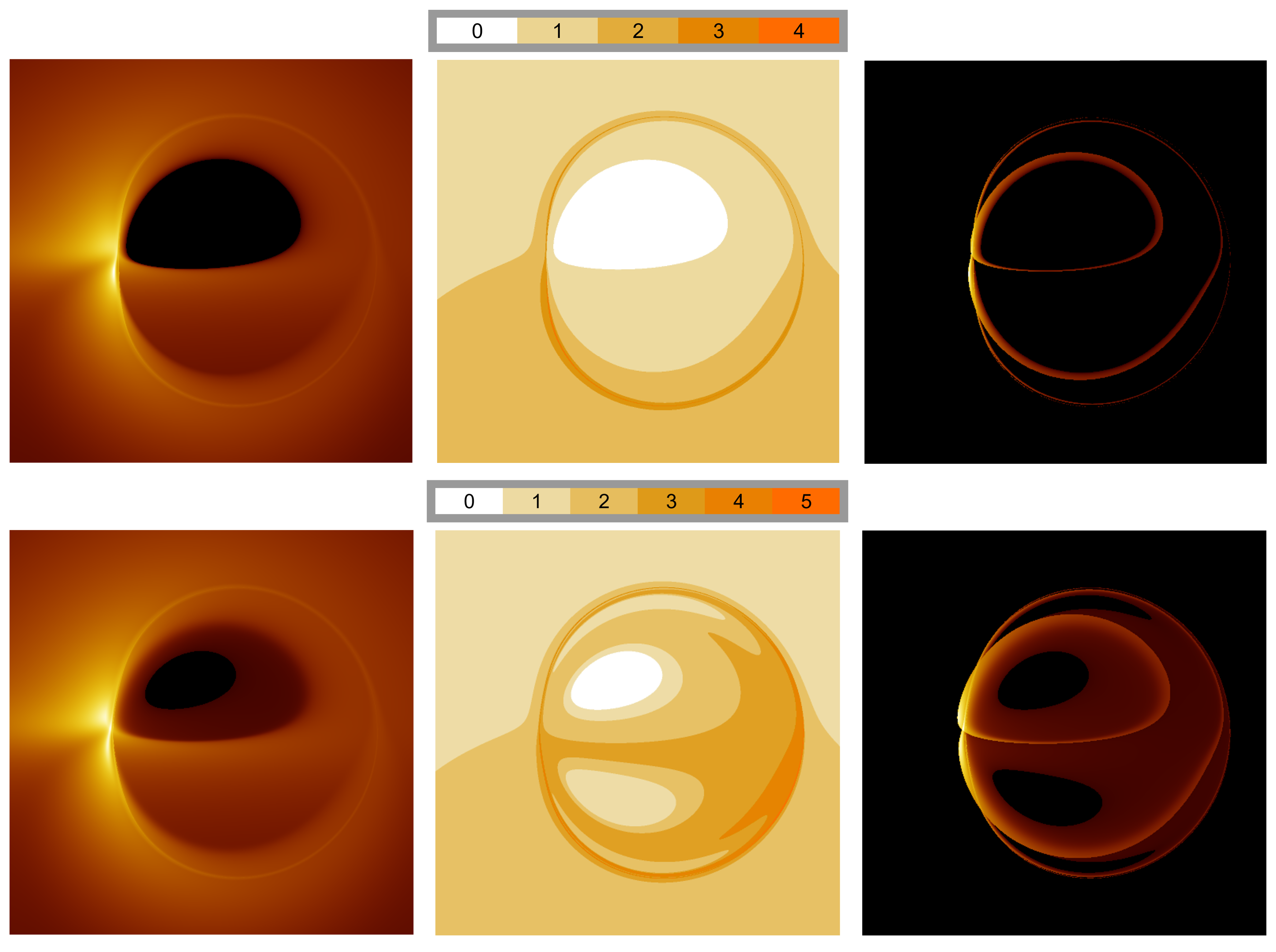}}
	\caption{Black hole images illuminated by the thin disk model. The spin is set to be $a=0.99$, and the observer is positioned at $r_o=200$, $\theta_o=80^\circ$. The top row corresponds to a typical Kerr black hole with $\alpha=0$, while the bottom row represents the quantum-corrected Kerr black hole with $\alpha=0.02$.  In the left column, we display the intensity maps generated by the thin disk. The middle column shows the number of times the light rays intersect the disk. Finally, the right column exhibits the intensity maps generated solely by the inner-ISCO region of the disk.}
	\label{disk}
\end{figure}

The left column of Figure~\ref{disk} illustrates the intensity maps obtained from the thin disk imaging. In this case, the spin parameter is fixed at $a=0.99$, and the observer is situated at $r_o=200$, $\theta_o=80^\circ$. Within these intensity maps, the bright, narrow curves are referred to as ``photon rings", which emerge due to the strong gravitational lensing effect. The shapes of these photon rings align with the shadows cast by the celestial light source mentioned earlier. Moreover, the completely dark regions central in the images result from the direct imaging of horizons, known as the ``inner shadows" \cite{Chael:2021rjo}. 

When comparing the two images in the left column of Figure~\ref{disk}, the most prominent distinction is that the quantum-corrected black hole exhibits a smaller inner shadow. To make a further investigation, we plot the maximal number of times light rays crossed the disk, $N_{max}$, as depicted in the middle column of Figure~\ref{disk}. Two notable differences are observed: First, the area occupied by light rays that do not intersect the disk is smaller in the image of a quantum-corrected black hole. As light rays accumulate intensity while passing through the disk, the inner shadow of a quantum-corrected black hole naturally appears diminished. 
	
Second, for the quantum-corrected black hole, the regions corresponding to $N_{max} = 3$ and $N_{max} = 4$ have extended ``tentacles" outward, which suggests that light rays are more likely to oscillate in their $\t$-direction near the equatorial plane of the quantum-corrected black hole. We have analyzed the trajectories of light rays associated with the tentacles and observed that they predominantly intersect the accretion disk between the ISCO and the horizon. This suggests that the non-circular spacetime effects resulting from quantum corrections are more pronounced in the vicinity of the horizon. In other words, the contribution of a near-horizon light source plays a crucial role in the observable effects of a quantum-corrected black hole. To verify this, we present the imaging results in the right column of Fig.~\ref{disk}, where only the light source inside the ISCO is present. It is evident that the image of the quantum-corrected black hole differs significantly from that of a typical Kerr black hole. We would like to stress that though the conventional disk model considers the ISCO as its inner boundary, there are also studies indicating that the accretion flow within the ISCO exhibits nontrivial dynamics and significant radiative capabilities \cite{Mummery:2023tgh}. Therefore, we may expect the presence of observable effects of quantum corrections in realistic accretion systems of black holes \footnote{It is worth noting that we have employed a phenomenological formula \eqref{ems} for the emissivity when generating the images. Realistic millimeter-wave emissions are dominated by thermal or non-thermal synchrotron radiations, which are determined by the thermodynamics of the accretion flow.}.

\section{Summary and discussion} \label{sec5}

In this work, we studied the LRs, shadows and disk images of the Kerr black hole in the semiclassical gravity with type-A trace anomaly \cite{Fernandes:2023vux}. Considering that the mass function $\mM$ depends both on $r$ and $\t$, the spacetime should be non-circular. Consequently, the Hamiltonian for geodesics does not allow for the separation of variables. To determine the radii of LRs, we performed numerical calculations based on the effective potential of photons. Our results demonstrated that the radius of the retrograde LR are always larger than that of the prograde LR, regardless of the value of the coupling constant. 

Furthermore, we employed a celestial sphere to illuminate the black hole and utilized the backward ray-tracing technique to determine the shape and area of the shadow on the observer's screen. To characterize the variations in the shadow's area, we introduced the parameter $\eta=S_{\text{BH}}/S_{\text{Kerr}}$, representing the ratio of the shadow's area for a quantum-corrected black hole to that of a Kerr black hole with the same spin. Our findings revealed that as the coupling constant gradually increased within its range of values, $\eta$ also increased. Moreover, an intriguing discovery was that the NHEKline was highly susceptible to disruption by the coupling constant. This phenomenon may arise from the fact that the presence of the trace anomaly breaks the degeneracy between the inner and outer horizons, resulting in the absence of the NHEK geometry.

After that, we employed a thin disk model to study the black hole images in a more realistic manner. Our findings revealed that the images of the quantum-corrected black hole could deviate from those of the typical Kerr black hole. The quantum correction leads to a higher frequency of light ray oscillations around the equatorial plane, specifically between the ISCO and the horizon. As a result, the influence of a source near the horizon becomes crucial in determining the observable effects of a quantum-corrected black hole.

\section*{Acknowledgments}
We thank Prof. Bin Chen for the valuable discussions. The work is partly supported by NSFC Grant No. 12275004, 12205013 and 11873044. MG is also endorsed by ”the Fundamental Research Funds for the Central Universities” with Grant No. 2021NTST13.

\appendix

\bibliographystyle{utphys}
\bibliography{anomalyBH}

\end{document}